# Optimization and Neural Network-Based Modelling of Surface Passivation Effectiveness by Hydrogenated Amorphous Silicon for Solar Cell Applications

Rahul Goyal[1], Sachin Kumar[2]

[1]Centre for Nanoscience and Engineering, Indian Institute of Science, Bangalore, India, email: rahulgoyal@iisc.ac.in

[2]Department of Materials Engineering, Indian Institute of Science, Bangalore, India, email: ksachin@iisc.ac.in

*Abstract*— Intrinsic hydrogenated amorphous silicon films can provide outstanding surface passivation of crystalline silicon wafer surfaces. This quality of Intrinsic hydrogenated amorphous silicon makes it valuable in heterojunction with intrinsic thin layer (HIT) solar cell fabrication. This paper describes the material characteristics and electronic properties of Intrinsic hydrogenated amorphous silicon that affects its passivation quality. A study of passivation quality of intrinsic hydrogenated amorphous silicon layer has been done with respect to deposition parameters in Plasma Enhanced Chemical Vapor Deposition (PECVD), the most commonly used method of its deposition. It was found that very good surface passivation with surface recombination velocity < 50 cm/s can be obtained from thickness of 30 nm of Intrinsic hydrogenated amorphous silicon (a-Si:H(i)), which is better than most other passivation techniques. A mathematical model based on Artificial Neural Network (ANN) is designed to predict the carrier lifetime for a given deposition condition and it is shown that the prediction capability of developed ANN model varies with the number of neurons in the hidden layer using Akaike Information Criterion (AIC), which is a widely accepted model selection method for measuring the validity of nonlinear models.

## I. Introduction

Silicon photovoltaics technology can provide the world with a renewable, reliable and economical source of energy. Among the semiconductor materials with suitable optoelectronic properties for photovoltaic applications, silicon is the most widely used. Advantages of silicon are its abundance in nature, and a well-developed fabrication technology that can contribute in the development of highly efficient futuristic solar cells. Conventional solar cells, made of crystalline silicon, have remained the most common type till now. Although crystalline silicon solar cells technology has high photovoltaic conversion efficiency, it is also associated with high production cost. In conventional solar cells manufacturing, the cost of crystalline silicon wafer (thickness > 220 μm) is about 40% to 60% of the solar cell cost [1]. In order to reduce the cost of solar cells, a photovoltaics technology that makes more efficient use of the expensive silicon material is required. Silicon heterojunction (SHJ) solar cell technology provides one such alternative; it is a combination of both crystalline and amorphous silicon technology in the form of a heterojunction. A heterojunction with intrinsic thin layer (HIT) solar cell can provide high performance together with the possibility of reducing silicon wafer thickness below 100 μm [2].

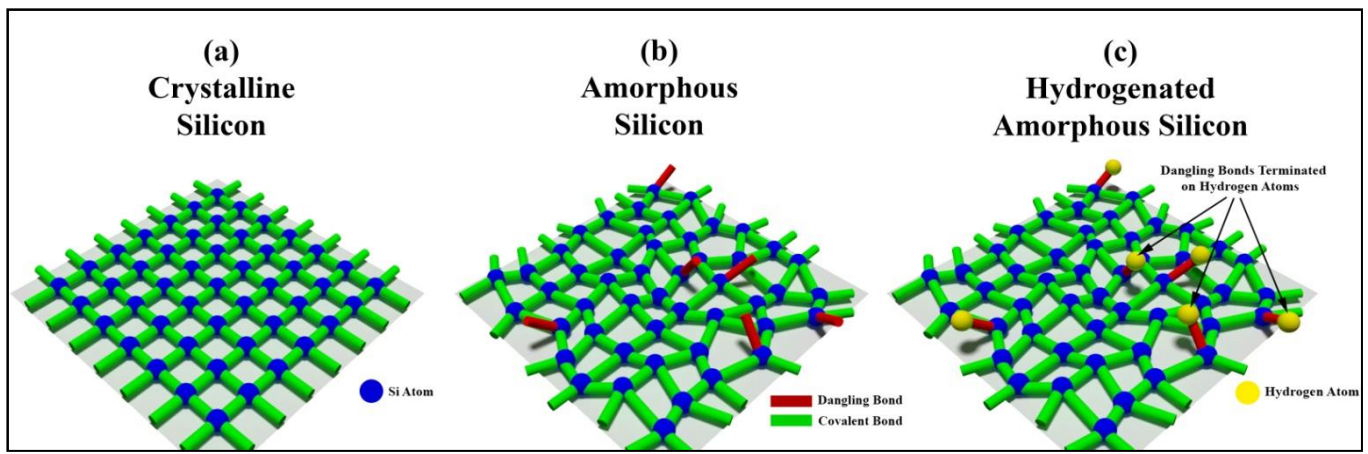

Fig. 1. Two-dimensional schematical represenatation of bonding configuration in (a) Crystalline Silicon, (b) Amorphous Silicon, and (c) Hydrogenated Amorphous Silicon.

A HIT solar cell is composed of a thin crystalline silicon (c-Si) wafer surrounded by ultra-thin amorphous silicon layers, then a intrinsic hydrogenated amorphous silicon (a-Si:H) passivation layer followed by a p/n-doped amorphous silicon layer, both deposited by plasma enhanced chemical vapor deposition (PECVD). On top of the silicon layers, an antireflective transparent conductive oxide (TCO) like indium tin oxide (ITO) is deposited by physical vapor deposition (PVD) and the charge collection is made by a screen-printed metallic contacting grid. The four principal areas of research around silicon-based heterojunction solar cells are: a) Surface passivation, b) Contact formation, c) Device architecture, and d) Operating conditions. In this paper our focus is on surface passivation provided by intrinsic a-Si:H passivation layer. A mathematical model based on Artificial Neural Network (ANN) is designed where modeling of surface passivation effectiveness of deposited intrinsic (a-Si:H) for various deposition parameters is done.

State of the art intrinsic a-Si:H layers are deposited using Radio Frequency (RF) Plasma-enhanced Chemical Vapor Deposition (PECVD), in which the deposition conditions crucially influence the surface passivation effectiveness. Here, the passivation efficiency of intrinsic a-Si:H layers deposited using PECVD at 13.56 MHz is quantified using carrier lifetime, which is compared at various deposition pressures, dilution ratios of precursors, and temperatures. In general, the systematic experimental approach to optimize the parameters involved in the passivation layer is not only time consuming but also requires special attention in cases where the contributions of parameters are correlated. Therefore, a good model can be of significant interest to simulate and predict the carrier lifetime from the variables involved in the deposition process. Several approaches are present to study the behavior of input parameters towards the output response properties but Artificial Neural Network (ANN) has been used to study this complex phenomenon as it is the most powerful technique to effectively analyze the effects of several independent variables simultaneously without the objective function between variables.

## II. EXPERIMENTAL DETAILS

Intrinsic a-Si:H layers are deposited on one side of double-side polished 4" n-type 1-5 Ωcm Fz wafers (100) using a conventional capacitively-coupled parallel-plate (CCP) RF PECVD reactor with a power density of up to 33 mWcm$^{-2}$. The scanning electron microscopy images of the deposited intrinsic a-Si:H are shown in Figure 2 and the uniformity and quality of the passivation film is clearly evident from the cross-sectional image of Figure 2. Prior to deposition, the wafers are cleaned using standard RCA solutions for 10 minutes, followed by a 5% HF dip for 1 minute to remove oxides from one of the surface of silicon wafer, rinsed in DI water, and dried in $N_2$ atmosphere. The transfer time from the dryer to the deposition chamber is kept to a minimum and is typically under 1 minute. The deposition chamber was always cleaned by the combination of gases $CF_4$ and $N_2O$ with $N_2$ purge before the deposition to minimize the possibility of contamination [3, 7]. All depositions are performed for various deposition parameters and used silane ($SiH_4$) as the precursor gas. The following parameters are varied: (a) deposition pressure in the range of 200-600 mTorr; (b) $H_2$ dilution ratio, defined as the ratio between $H_2$ and silane gas flow, $H_2/SiH_4$, in the range of 75-150%; (c) substrate temperature in the range of 200-350ºC. All samples undergo a 45-minutes dielectric anneal in a $N_2$ environment before deposition for measuring the bulk lifetime of the sample. The silicon wafer is first oxidized and then a thickness of 100 nm of oxide is grown for the measurement of bulk lifetime which is came out to be around 156 μs.

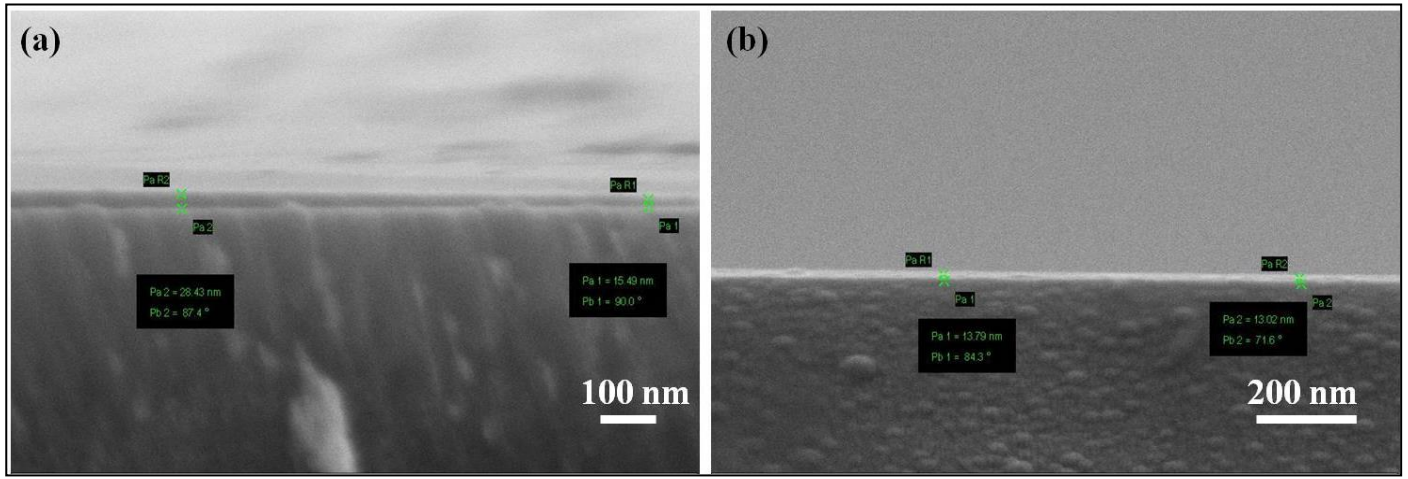

Fig. 2. Cross-sectional scanning electron microscopy images of intrinsic a-Si:H deposited on silicon wafers corresponding to the thickness of (a) 30 nm, (b) 15 nm.

After intrinsic a-Si:H layers are deposited, its thickness is measured using Spectroscopic Ellipsometry, which is later cross-checked using Scanning Electron Microscopy measurements. The effective lifetime is measured using a MDP Spot Freiberg Instruments lifetime tester in the transient mode at the wavelength of 980 nm. The surface lifetime is extracted at an excess carrier density of $2 \times 10^{16}$ cm$^{-3}$. The surface lifetime is calculated using the following equation:

$$\frac{1}{T_{EC}} = \frac{1}{T_{BC}} + \frac{1}{T_{SC}} \quad \ldots\ldots \text{Equation 1}$$

where $T_{EC}$ is effective carrier lifetime, $T_{BC}$ is bulk carrier lifetime and $T_{SC}$ is surface carrier lifetime.

## III. RESULTS AND DISCUSSION

In this section we review the deposition parameters varied in the experiment like pressure, temperature, dilution ratio, and thickness that affect the passivation quality of hydrogenated amorphous silicon [9, 10, and 11]. The quality of the deposited film is measured using the effective lifetime of the charge carriers. As the lifetime also depends on the thickness of the passivation layers, therefore variation of lifetime is measured for different values pressure, temperature, dilution ratio, while keeping the thickness constant. The note below provides a brief description on the effect of changes various deposition parameters and their range used in the experiment.

### A. Thickness (τ)

The hydrogenated amorphous silicon layer acts as a reservoir of hydrogen for passivation of crystalline silicon surface. The size of this reservoir increases with increase in thickness [4, 5, and 6]. Hence with increase in thickness, the passivation of the surface improves so does the effective carrier lifetime. The variation of the effective carrier lifetime with thickness is shown in Figure 3 and the saturation behavior of the lifetime variation can be clearly observed from the presented result. The maximum limit of effective carrier lifetime is equivalent to bulk carrier lifetime which is equal to 156 μs and at his point the surface carrier lifetime would be at its limiting value of infinity which demonstrates the perfect conversion of dangling bonds to covalent bonds. This variation of carrier lifetime as a function of thickness can be best explained from the Figure 1. The increase in thickness of the passivation layer will lead to the increase in the concentration of hydrogen radicals on the surface which will contribute in the termination of dangling bonds on the surface of the interfacing layer. Hence, the generated carriers will not be utilized for the completion of bonding configuration in silicon atoms; instead these carriers will play a significant role in the conductivity of solar cells. With the demonstration of experimental data in Figure 2, it also presents a second order polynomial fit for both the effective and surface carrier lifetime to mathematically understand the order of dependency. The equations for the second-order polynomial fit can be written as follows,

$$T_{EC} = -0.12\tau^2 + 7.57\tau + 7.18 \quad \ldots\ldots \text{Equation 2}$$

$$T_{SC} = -0.62\tau^2 + 6.91\tau + 7.82 \quad \ldots\ldots \text{Equation 3}$$

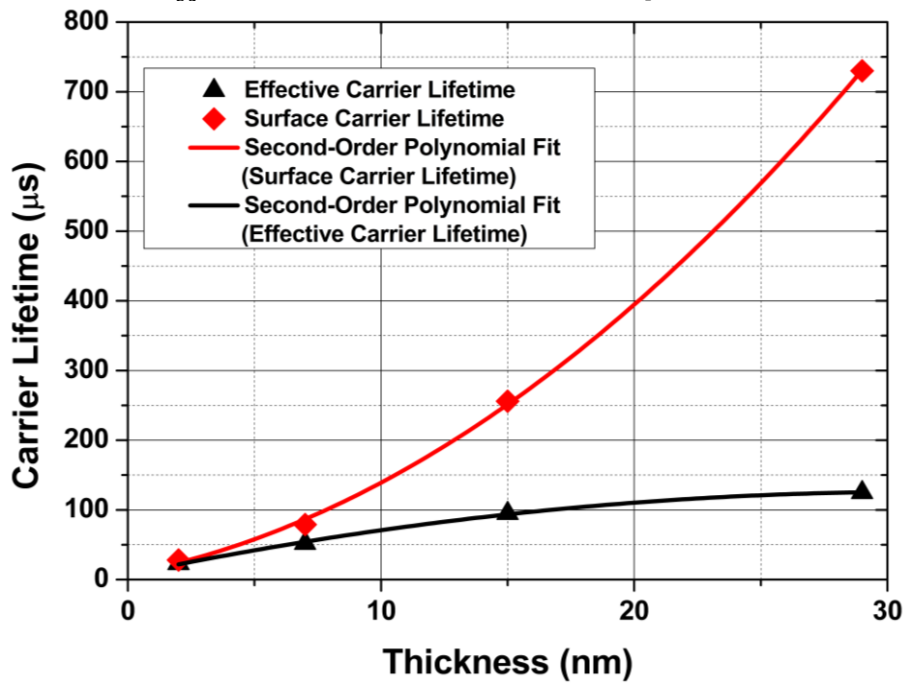

Fig. 3. Variation of carrier lifetime with change in depostion thickness

As the lifetime increases with increase in thickness, the study of other parameters of deposition was performed keeping the deposition thickness as constant.

### B. Deposition Pressure (P)

The gas pressure determines the mean free path for collisions of the gas molecules and influences whether the reactions are at the growing surface or in the gas [4, 5, and 6]. Deposition pressure in the range of 250-600 mTorr is presented. The variation of the effective lifetime with deposition pressure is shown in Figure 4. The lifetime peaks around 450mTorr. It is argued that with increase in pressure more hydrogen can be incorporated into the a-Si:H network thus providing better passivation for the film. According to the relationship provided by Cody et al. [8], the $H_2$ content estimated at 250 mTorr is less than that of estimated at 400 mTorr, which supports the above argument. With higher hydrogen content, more hydrogen atoms are available to repair the dangling bonds, saturating recombination centres and thus increasing the effective lifetime. However, this trend soon gets reversed as more Si-$H_2$ bonds are formed and Si-$H_2$ is known to degrade the electronic property of a-Si:H, hence decrease the passivation quality of the amorphous silicon film. It can be deduced form this observation that Si-$H_2$ configuration is more stable as compared to a-Si:H when the concentration of hydrogen atom is significantly more than the optimum concentration to terminate the dangling bonds. The concentration of the hydrogen is directly proportional to the partial pressure of the hydrogen gas in the deposition chamber which is related to the deposition pressure of the hydrogenated amorphous silicon in the PECVD system. Thus, the hydrogen concentration increases with the deposition pressure, but the lifetime increases for a certain range of deposition pressure with maximum at optimum hydrogen concentration and then decreases due to the formation of Si-$H_2$ bonds. This variation of carrier lifetime is observed in both the surface and effective lifetime as evident from Figure 4. The equations for the second-order polynomial fit can be written as follows,

$$T_{EC} = -0.001P^2 + 1.30 - 141 \quad \ldots\ldots \text{Equation 4}$$

$$T_{SC} = -0.022P^2 + 18.41P + 2742.6 \quad \ldots\ldots \text{Equation 5}$$

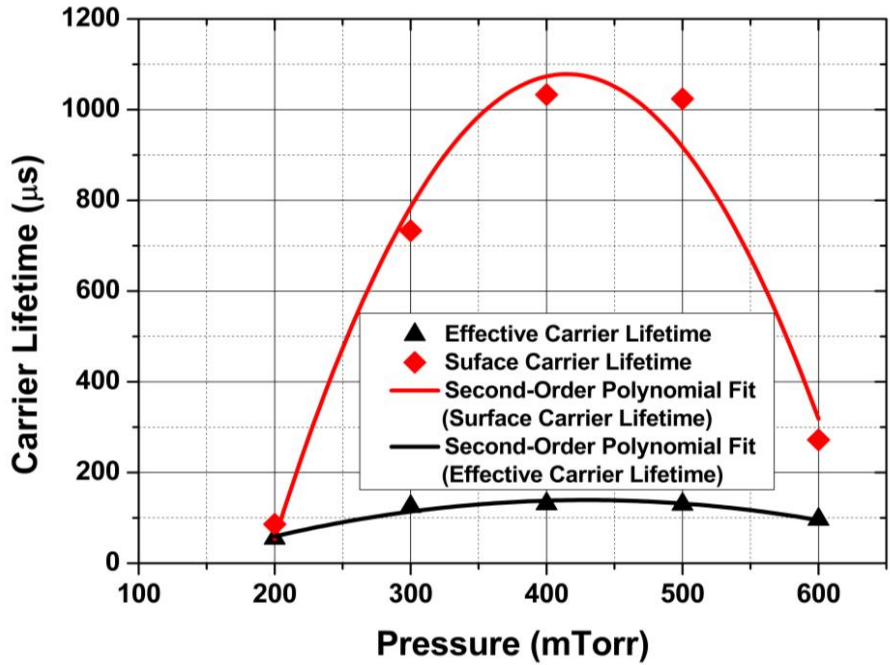

Fig. 4. Variation of carrier lifetime with change in pressure

C. *Hydrogen Dilution Ratio (χ)*

Hydrogen dilution ratio is the ratio of flow rate of hydrogen to the flow rate of Silane. The hydrogen dilution ratio determines the hydrogen concentration in the deposited film [4, 5, and 6]. This ratio is varied from 75% to 150%. The effective lifetime dependence on the hydrogen dilution ratio is shown in Figure 5. As the dilution ratio increases from 75 to 125%, the effective lifetime of the sample increases from 0.09 to 0.13 ms. However, with further increase in dilution ratio the lifetime decreases. The observed increase in lifetime is mainly due to the higher $H_2$ incorporation into the amorphous network. The lifetime decreases to significantly lower values beyond 100; this sharp decrease in the surface carrier lifetime is due to an important bonding-rearrangement behavior that can explain the variation of carrier lifetime with hydrogen dilution ratio. This happens because of the transition of amorphous silicon film to crystalline silicon film in the presence of increased $H_2$ dilution ratio. This conversion makes the amorphous silicon incapable to contribute in the termination of dangling bonds on the interface layer and leads to films with poor passivation quality. Therefore, it can be argued that the amorphous film deposited close to the transition state of amorphous to crystalline can provide the best passivation quality. The equations for the second-order polynomial fit can be written as follows,

$$T_{EC} = -300\chi^2 + 661\chi - 236.25 \quad \ldots\ldots \text{Equation 6}$$

$$T_{SC} = -3754\chi^2 + 8409.5\chi + 3979.17 \quad \ldots\ldots \text{Equation 7}$$

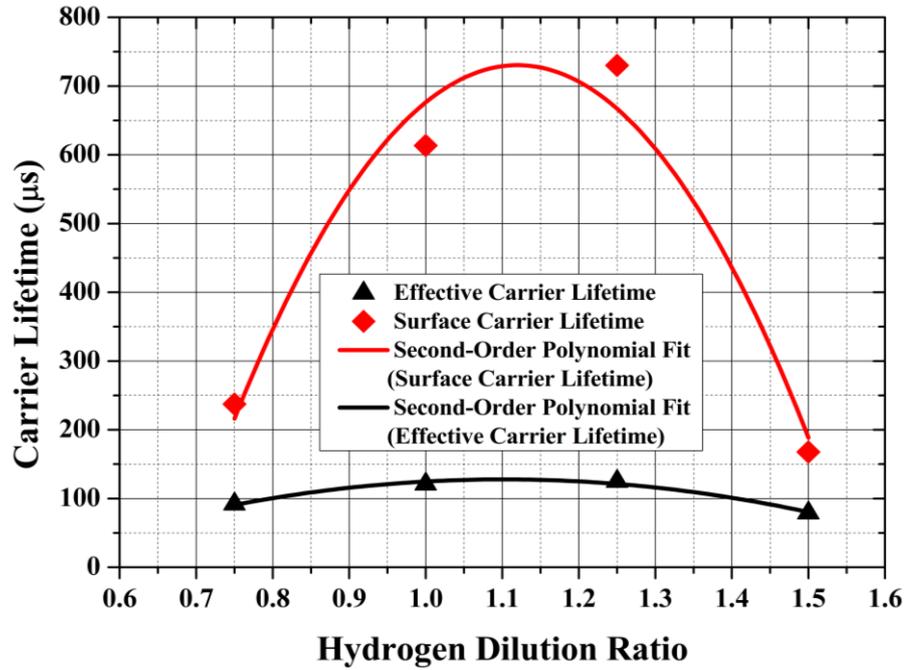

Fig. 5. Variation of carrier lifetime with change in Hydrogen Dilution

### D. Deposition Temperature (T)

The temperature of the substrate controls the chemical reactions on the growing surface. The Substrate temperature variation presented is in the range of 200 – 350 °C. Figure 6 shows the carrier lifetime variation with change in substrate temperature during deposition [4, 5, and 6]. The lifetime decreases from 0.11 ms to about 95 μs, when the substrate temperature increases from 200 °C to 350 °C. Higher temperature of deposition (>250°C) leads to the increase in the concentration of defects and decrease in the concentration of hydrogen hence responsible for creating more dangling bonds and thus decreases the quality of surface passivation. From the set of experiments performed, the optimum temperature of deposition is found to be around 200°C. The equations for the second-order polynomial fit can be written as follows,

$$T_{EC} = -0.0006\text{T}^2 - 0.202\text{T} - 207.3 \quad \ldots\ldots \text{Equation 8}$$

$$T_{SC} = 0.06\text{T}^2 - 46.93\text{T} + 8584 \quad \ldots\ldots \text{Equation 9}$$

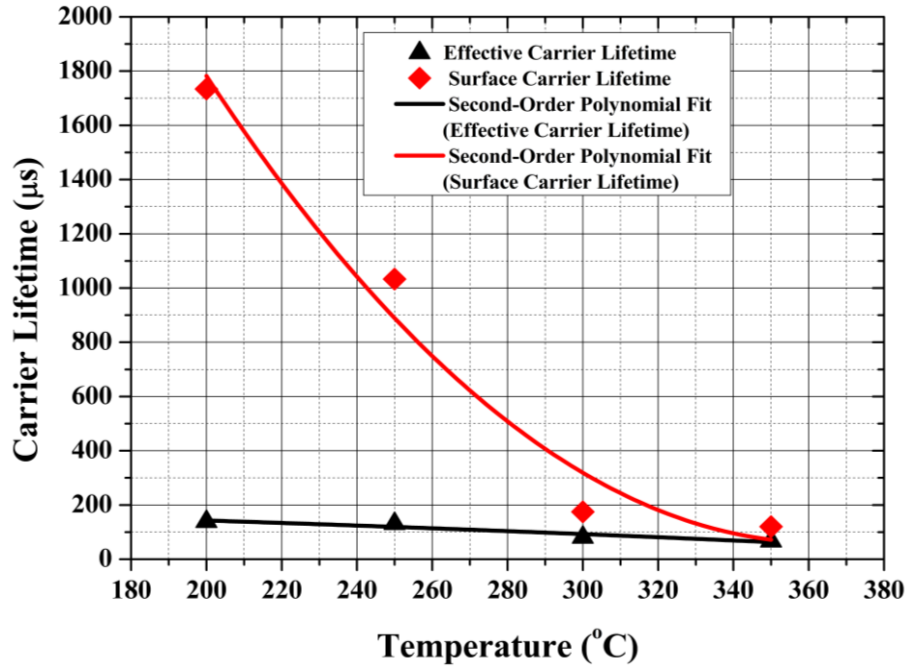

Fig. 6. Variation of carrier lifetime with change in Deposition Temperature

## IV. MODELLING OF CARRIER LIFETIME USING ARTIFICIAL NEURAL NETWORK

The dependency of four different input variables on the response of carrier lifetime (L) was studied by the ANN-based model, developed during the present study. It provides the modeling of complex nonlinear relationships for describing the output response of carrier lifetime. The first, second and third variable is the pressure, temperature and dilution ratio, respectively in the intrinsic a-Si:H layers' passivation system. The fourth variable is the thickness of the intrinsic a-Si:H layers.

An ANN-based model is basically inspired by the structural and functional aspect of biological neural network which acts as an interface between sensory organs and brain and has proved to be the most effective modeling method in recent years [12], particularly for process modeling. The structure of the model consists of an input layer which comprises of the independent variables, a number of hidden layers and an output layer basically denotes dependent variables as shown in Figure 7. The layers are interconnected through a number of processing units called neurons, these neurons along with the weighted connections or bias factor acts as a linkage between input layer, hidden layers and output layer. The function of hidden layer is to process data and produces an output based on the sum of mathematically optimized weighted values from the input layer modified by a sigmoid transfer function. The sigmoid transfer that is used in the current study is given by,

$$f(x) = -1 + \frac{2}{(1+ e^{-2x})} \quad \ldots\ldots \text{Equation 10}$$

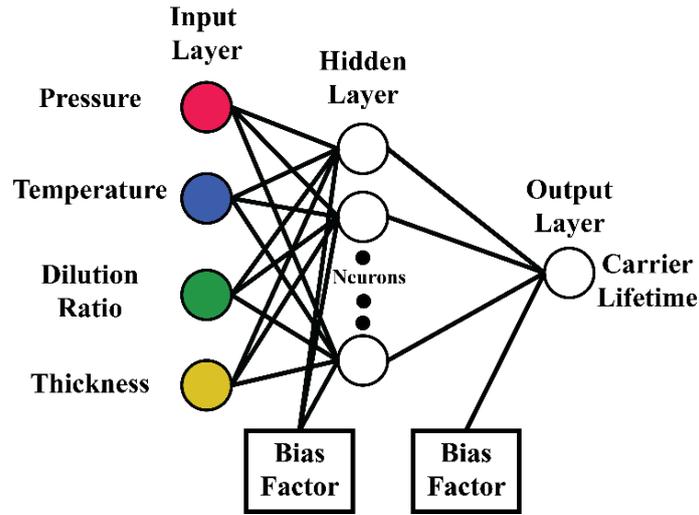

Fig. 7. General scheme of artificial neural network

The topology of the network starts with the efficient training of the neural network and for that Levenberg-Marquardt back-propagation algorithm is used for network training. The Neural Network Toolbox of MATLAB Version R2016a is used to optimize and perform calculations for designing the ANN-based mathematical model. Because, the network is developed to represent only one type of response i.e. carrier lifetime therefore, the topology of the ANN developed is designated as 4-$n_h$-1. The mathematical architecture of the Artificial Neural Network is shown in Figure 7. The network contains four input neurons representing the four variables for the deposition conditions i.e. pressure, dilution ratio, temperature and thickness, $n_h$ represents the number of hidden neurons in a single hidden layer, and one output neuron representing one type of output response. The experimental data is used to build, train, test and validate the ANN topologies for a number of hidden neurons varying from 1 to 30. The trial and error search method is used for the training process, and the program is run until an optimum model is found and the minimum of mean square error is reached in the validation process. The conventional use of $R^2$ method to measure the accuracy of fit in the nonlinear models is pointed out to be inefficient [13], to supplement this deficiency, Akaike Information Criterion (*AIC*) [14] is used to measure the validity of nonlinear modeling [15] of ANN.

$$AIC = 2K - 2\ln(m_L) \quad \ldots\ldots \text{Equation 11}$$

where $K$ = number of estimable parameters and $\ln(m_L)$ = empirical log-likelihood function of the estimated model at its maximum point. The maximum log-likelihood value for the case of non-linear fit can be estimated from:

$$\ln(m_L) = \frac{1}{2} \times \left(-N \times \left(\ln(2\pi) + 1 - \ln(N) + \ln(\sum_{i=1}^{N} r_i^2)\right)\right) \quad \ldots\ldots \text{Equation 12}$$

where $r_i$ denotes the residuals from the nonlinear least-squares fit and $N$ symbolizes their number. The empirical log-likelihood function can be approximated to a simplistic form by taking an assumption that, if

$$\ln\left(\sum_{i=1}^{N} r_i^2\right) \gg \ln(2\pi) + 1 - \ln(N) \quad \ldots\ldots \text{Equation 13}$$

then,

$$ln(m_L) = \frac{1}{2} \times \left(-N \times \left(ln\left(\sum_{i=1}^{N} r_i^2\right)\right)\right) \quad \ldots\ldots \text{Equation 14}$$

and this results to

$$AIC = 2K + \left(N \times \left(ln\left(\sum_{i=1}^{N} r_i^2\right)\right)\right) \quad \ldots\ldots \text{Equation 15}$$

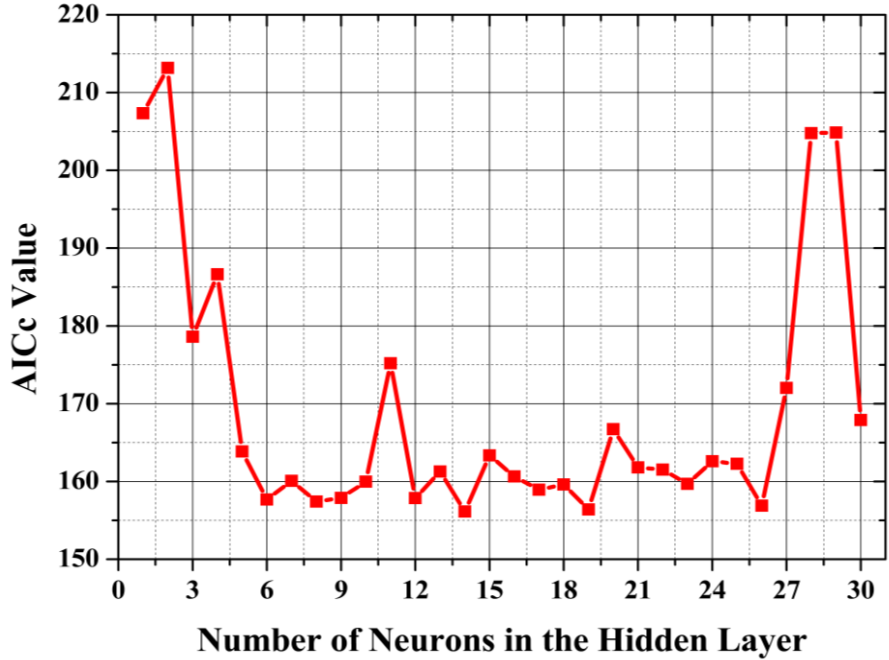

Fig. 8. Value of the bias-corrected Akaike Information Criterion as a function of number of nodes in the hidden layer of the neural network

The ANN-based model contains bias factors that create weightage in the error residuals and therefore provide inconsistent results. This mathematical biasing in the results of *AIC* can be removed by the utilization of bias-corrected AIC variant (*AICc*) which is defined as,

$$AICc = AIC + \frac{2K(K+1)}{n_s - K - 1} \quad \ldots\ldots \text{Equation 16}$$

where $n_s$ is the sample size. According to the operational principle of *AIC*, the model that corresponds to the minimum value of AICc denotes the best fit nonlinear model for the design experiment. The calculated AICc values for the developed ANN network for the carrier lifetime prediction is shown in Figure 8, and it is clearly evident that the number of hidden nodes that would be sufficient to build a efficient topology of the neural network is 14. The designed optimized neural network for 14 number of hidden nodes is graphically represented in Figure 9. The training and validation data sets are explicitly demonstrated in the model and the validity of the model is graphically interpreted using a straight line of unit slope and zero intercept. Thus, the modeling based on artificial neural networks can be used to design intelligent and intensive experimentation for controlling the complexity of the variables involved in the fabrication of solar cells.

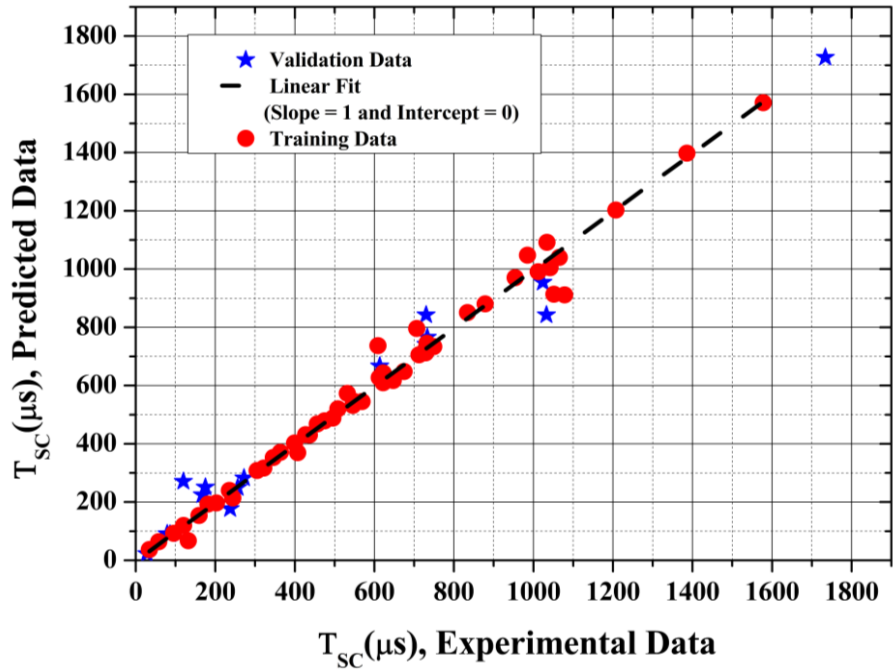

Fig. 9. Variation of carrier lifetime with change in Deposition Temperature

V. CONCLUSION

Surface defects of crystalline silicon and its passivation using intrinsic a-Si:H, and the optimum conditions for deposition of good quality intrinsic a-Si:H was discussed. Varying the deposition pressure, $H_2$ dilution, and the substrate temperature, key performance indicator, i.e., surface lifetime, was compared and analyzed. An artificial neural network-based model was also built which can be used for predicting the carrier lifetime for various deposition parameters. Further, the following points can be concluded from the experiments:

- The effective lifetime for intrinsic a-Si:H thin film is a function of the total $H_2$ present in the amorphous network. In particular, the best passivation quality results from films with a high Si-H bond concentration and a low Si-$H_2$ concentration.
- The optimal film can be obtained by tuning the deposition pressure and the $H_2$ dilution ratio.
- Higher temperature of deposition (>250°C) leads to higher defects and lower hydrogen concentration thus poor passivation.

ACKNOWLEDGMENT

This work was carried out at National Nanofabrication Center (NNFC) and Micro and Nano Characterization Facility (MNCF) located at Centre for Nanoscience and Engineering (CeNSE), Indian Institute of Science, Bangalore. We would like to thank Professor Sushobhan Avasthi for his guidance and Nizam Subhani for helping with the PECVD experimental setup.